%% file: MAIN.tex
\documentclass[16pt, twoside]{pnas-new}
\setboolean{displaywatermark}{false}
\DeclareUnicodeCharacter{202A}{\,}

\templatetype{pnasresearcharticle} 
\usepackage{setspace} 
\usepackage{float}

\usepackage{graphicx}
\usepackage{dcolumn}
\usepackage{bm}
\usepackage{siunitx}

\nolinenumbers

\usepackage{amsmath}
\usepackage[colorlinks=true, allcolors=blue]{hyperref}
\usepackage[T1]{fontenc}
\usepackage{subfiles} 
\usepackage{xcolor}
\usepackage{easyReview}

\begin{abstract}
Addition of particles to a viscoelastic suspension dramatically alters the properties of the mixture, particularly when it is sheared or otherwise processed.  Shear-induced stretching of the polymers results in elastic stress that causes a substantial increase in measured viscosity with increasing shear, and an attractive interaction between particles, leading to their chaining.  At even higher shear rates, the flow becomes unstable, even in the absence of particles. This instability makes it very difficult to determine the properties of a particle suspension.  Here we use a fully immersed parallel plate geometry to measure the high-shear-rate behavior of a suspension of particles in a viscoelastic fluid.  We find an unexpected separation of the particles within the suspension resulting in the formation of a layer of particles in the center of the cell.  Remarkably, monodisperse particles form a crystalline layer which dramatically alters the shear instability.  By combining measurements of the velocity field and torque fluctuations, we show that this solid layer disrupts the flow instability and introduces a new, single-frequency component to the torque fluctuations that reflects a dominant velocity pattern in the flow. These results highlight the interplay between particles and a suspending viscoelastic fluid at very high shear rates.

\end{abstract}
\begin{document}

\title{Anomalous crystalline ordering of particles in a viscoelastic fluid under high shear}

\author[a]{Sijie Sun}
\author[b,c]{Nan Xue} 
\author[a,d]{Stefano Aime}
\author[a,f]{‪Hyoungsoo Kim}
\author[a,g]{Jizhou Tang}
\author[e]{Gareth H. McKinley}
\author[b]{Howard A. Stone}
\author[a,1]{David A. Weitz}
\affil[a]{John A. Paulson School of Engineering and Applied Sciences, Harvard University, Cambridge, Massachusetts, 02138, USA}

\affil[b]{Department of Mechanical and Aerospace Engineering, Princeton University, Princeton, New Jersey 08544, USA}
\affil[c]{Department of Materials, ETH Z{\"u}rich, Z{\"u}rich, 8093, Switzerland}
\affil[d]{ESPCI, Paris, 75005, France}
\affil[e]{Department of Mechanical Engineering, Massachusetts Institute of Technology, Cambridge, Massachusetts, 02139, USA }
\affil[f]{Department of Mechanical Engineering, Korea Advanced Institute of Science and Technology (KAIST), 291 Daehak-ro, Yuseong-gu, Daejeon 34141, Republic of Korea}
\affil[g]{State Key Laboratory of Marine Geology, Tongji University, Shanghai,201804, China}
\correspondingauthor{\textsuperscript{1}To whom correspondence should be addressed. E-mail: weitz@seas.harvard.edu}

\date{\today}

 \maketitle

\ifthenelse{\boolean{shortarticle}}{\ifthenelse{\boolean{singlecolumn}}{\abscontentformatted}{\abscontent}}{}

Suspensions of solid particles in a fluid are widely encountered in all ranges of technology both as products and as precursor materials in manufacturing and materials preparation. They exhibit a wide range of properties depending on the particle concentration or volume fraction, $\phi$, and on the nature of the suspending fluid.  When the suspending fluid is Newtonian, the viscosity of the suspension increases slowly with increasing $\phi$ at low volume fractions, and then diverges as $\phi$ approaches the maximum volume fraction of randomly packed solid spheres~\cite{wyart2014discontinuous, Weeks2014,brady_bossis_1985}.  As the shear rate increases, higher-volume-fraction suspensions exhibit slight shear thinning with the viscosity decreasing with increasing shear rate, while at very high shear rates, they undergo a sudden, dramatic shear thickening, becoming almost solid-like due to increased interparticle collisions that result in frictional interaction between the particles~\cite{seto2013discontinuous, Melrose1996}.  The behavior of a suspension is significantly different if the suspending fluid is viscoelastic~\cite{varga2019hydrodynamics,tanner2019rheology,d2015particle,shaqfeh2019rheology,shaqfeh2021oldroyd}.  At low $\phi$ and low shear rates, the particles follow streamlines, and the behavior of the suspension is similar to that of the suspending fluid~\cite{barnes_review_2003,mewis2012colloidal}. As the shear rate increases, however, the large polymer molecules become increasingly stretched and cannot fully relax, resulting in a stress normal to the streamline, causing particles to cross streamlines~\cite{huang1997direct,villone2013particle,liu1993anomalous,johnson1990dynamics,d2010viscoelasticity}, particle chaining~\cite{michele1977alignment,feng1996motion,won2004alignment,scirocco_effect_2004,xie2016flow,pasquino2010directed,vermant2005flow} and shear thickening~\cite{vazquez2019shear,yang_mechanism_2018,matsuoka_prediction_2021,zaccone2014linking}.  At high shear rates, the stretched polymer exerts yet larger elastic stresses; even in the absence of particles, these stresses dominate over viscous stress and can destabilize the flow field, resulting in an elastic instability and so lead to the development of a secondary chaotic flow~\cite{groisman_elastic_2004,mckinley_observations_1991,shaqfeh1996purely, datta2022perspectives,walkama2020disorder,magda1988transition,schiamberg2006transitional}. This instability makes it challenging to quantify the behavior of a viscoelastic suspension of particles at high shear rates; as a result, only low shear-rate regimes, where the instability is avoided, have been investigated, and the behavior of particles at high shear rates remains unexplored. However, the interplay between fluid elasticity and particles at high shear rates is important for a wide range of industrial processes, such as hydraulic fracturing~\cite{goel2002correlating,barbati2016complex,browne_shih_datta_2020} and polymer composite processing~\cite{astrom2018manufacturing,barnes2003review}, as well as natural phenomena like biofluid flow~\cite{Brust2013,stokes2007viscoelasticity,kumar2012mechanism}. 

In this paper, we explore the interactions between particles and a viscoelastic fluid at high shear rates.  We study flow between parallel plates in a rheometer and discover a surprising crystalline structure of particles that forms in the middle layer of the flow cell. This structure suppresses the elastic instability and significantly alters the flow dynamics.  In the absence of particles, the elastic instability leads to strong fluctuations in the torque measured with the rheometer. These fluctuations have a power-law spectrum reminiscent of turbulent flow.  The particles result in an emergence of a single dominant frequency in the fluctuations that persists above the underlying power-law spectrum. Using flow visualization techniques, we show that this single frequency is a result of rigid body rotation of a nearly static structure formed by the crystalline layer of particles.  These measurements highlight the complex interactions between particles and a viscoelastic background fluid at high shear rates.

\section*{Influence of the particles on the viscoelastic secondary flow}
The particles are monodisperse Poly(methyl methacrylate) (PMMA) with a diameter of 51 {\textmu}m (CA50 from MICROBEBS). The solution phase is a mixture of 84.596 wt\% 2-2 Thiodiethanol (CAS 111-48-8 from Sigma) and 15.404 wt\% deionized water. This mixture has a refractive index of 1.488 at 20°C, which precisely matches that of the PMMA particles. The density of the solvent is 1.18 g/ml, which is close to that of the particles (1.19 g/ml). We add 0.25 wt\% polyethylene oxide (PEO, CAS 25322-68-3 from Sigma) of molecular weight 8M Da to the solvent to make it viscoelastic. At a shear rate of $10~\mathrm{s}^{-1}$, the ratio of elastic to viscous stress is approximately 6.

\begin{figure*}[h]
\centering
\includegraphics[width=7in]{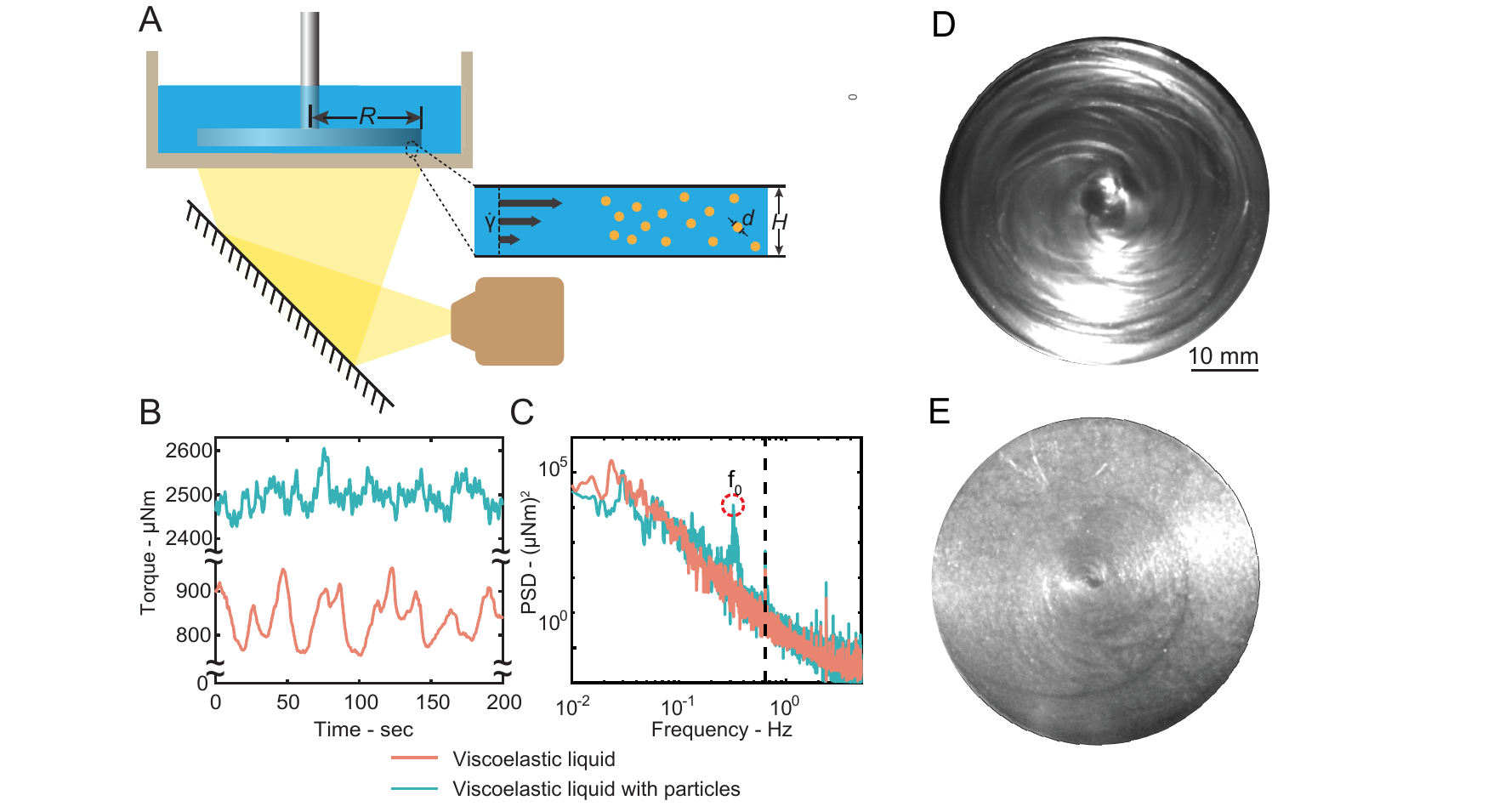}
\caption{Elastic instability in a viscoelastic fluid with and without  suspended particles. ({\it A}) Sketch of the setup. Optics for imaging is mounted beneath a rheometer. Viscoelastic liquid with and without monodisperse particles are sheared between parallel plates. A cup is added to the bottom plate to avoid spilling the sample under high shear rates. ({\it B}) The  torque reading $M$ measured by the rheometer; fluctuations are evident. The gap size $H = 2$ mm and the radius of the upper plate $R = 25$ mm. The jade green and coral orange lines show the torque fluctuations for the experimental conditions described in ({\it A}), with and without 30 vol\% particles ($d = 51$ {\textmu}m), respectively. The shear rate $\dot{\gamma} = 50~\mathrm{s
^{-1}}$. ({\it C}) The Power Spectral Density (PSD) of the torque measurement in ({\it B}). {The PSD is calculated based on 580 s torque measurement}. The PSD of the polymer solution has a power-law relationship with the frequency, implying the flow is turbulent-like. With the addition of the suspension, the torque has a single peak with a frequency of $f_0 \approx 0.323$ Hz, highlighted by the red circle. The single peak disrupts the original power-law decay. The black dashed line represents the rotation frequency of the upper plate. A five-point moving average is applied to the PSD to highlight the trend.
({\it D}) The flow pattern of the viscoelastic liquid without suspension. $H = 2$ mm, $R = 25$ mm, and $\dot{\gamma} = 50~\mathrm{s
^{-1}}$. Multiple spiral patterns coexist. ({\it E}) The secondary flow pattern with 30 vol\% particles (diameter $d = 51$ {\textmu}m). The other experimental conditions are the same as ({\it D}). One spiral pattern dominates over the others.
}
\label{fig:1}
\end{figure*}
We use a stress-controlled rheometer (MCR 501 from Anton Paar) with a parallel plate geometry that has a 50 mm diameter and a 2 mm gap. At high shear rates, the viscoelastic fluid can entrap air bubbles and expel liquid from the flow cell, making measurements difficult~\cite{keentok1999edge}. To overcome this problem, we modify the rheometer geometry by adding a cup to the lower plate {and load a large volume of sample to the rheometer}. This allows the rotating upper plate to be fully immersed in the fluid, as shown in Fig.~\ref{fig:1}{\it A}. This eliminates the liquid-air interface at the edge of the plates, thereby preventing the expulsion of the sample from between the parallel plates. We add enough fluid to ensure that the velocity gradient is much larger between the plates than between the upper surface and the free surface on the top.

In the absence of particles, the torque is constant and independent of time at low shear rates. As the shear rate increases, the solvent exhibits shear thinning, but the torque remains time independent. With further increase of shear rate, the solvent undergoes a sudden and dramatic shear thickening, and the torque exhibits pronounced fluctuations in time. We characterize the magnitude of these fluctuations by the ratio of the standard deviation of the torque ${\sigma_ M} $ to the mean value of the torque ${\overline{M}}$. For a shear rate of $50~\mathrm{s}^{-1}$, the fluctuations have a relative magnitude of 6\%, as shown in Fig.~\ref{fig:1}{\it B}. The torque fluctuations suggest that the flow is time-dependent and unstable, reflecting the presence of a secondary flow. To further characterize these fluctuations, we calculate the Fourier transform of the torque and determine the Power Spectral Density (PSD) from the square of the amplitude of the fluctuations as a function of frequency. The PSD exhibits a power-law decay proportional to $  f^{-k}$, where $k \approx 2.8$ for frequencies, $f$, above a rollover frequency of approximately $0.02$ Hz, as shown in Fig.~\ref{fig:1}{\it C}. The power law spectrum is reminiscent of a chaotic flow.

Upon adding particles at a volume fraction of 30\%, the suspension exhibits shear thinning as the shear rate increases similar to the behavior in the absence of the particles, as shown in Figure~\ref{fig:A8} in the SI. At even higher shear rates, the particle suspension exhibits a pronounced onset of shear thickening similar to the behavior in the absence of the particles; interestingly, however, this occurs at a lower shear rate and the degree of shear thickening is not as large as it is in the absence of particles.
The magnitude of the torque is considerably larger at a shear rate of 50~{$\hbox{s}^{-1}$}; the time dependence again exhibits fluctuations but with a single, dominant frequency, as shown in Fig.~\ref{fig:1}{\it B}. This response is reflected in the PSD, which still exhibits a power-law decay proportional to $  f^{-k}$, where $k \approx 2.8$. However, at a frequency of $f_0 = 0.323$ Hz, the PSD also exhibits a sharp peak that is two orders of magnitude larger, as shown by the red dashed circle in Fig.~\ref{fig:1}{\it C}. The presence of a pronounced peak in the PSD spectrum suggests that the suspended particles modify the chaotic flow.

To visualize the secondary flow, we employ a transparent bottom plate and add titanium-dioxide-coated mica flakes (0.5 wt\% from HTVRONT) to form a rheoscopic liquid. Under shear, the mica flakes align with the flow direction, allowing visualization of the secondary flow. For the polymer solution alone, the onset of the secondary flow is observed at the same shear rate that fluctuations in torque occur. At higher shear rates, when the elastic instability is fully developed, we observed multiple outward-flowing spirals that intermingle, but there is no dominant pattern, as shown for a shear rate of $50~\mathrm{s}^{-1}$ in Fig.~\ref{fig:1}{\it D} and Movie S1.

The introduction of particles leads to a significant alteration in the patterns of secondary flow resulting from the elastic instability. The multiple intermingled spiral patterns are no longer discernible. Instead, a single dominant spiral propagates from near the center to the periphery, as depicted in Fig.~\ref{fig:1}{\it E}. and Movie S2. The frequency of the spiral rotation is 0.3 Hz, which closely corresponds to the dominant frequency of the torque fluctuations, 0.32 Hz. This observation suggests that the primary, single-frequency torque oscillation stems from the pronounced spiral in the flow pattern.

\section*{ Dynamic Mode Decomposition of the flow structures}

\begin{figure*}[h]
\centering
\includegraphics[width=4.5 in]{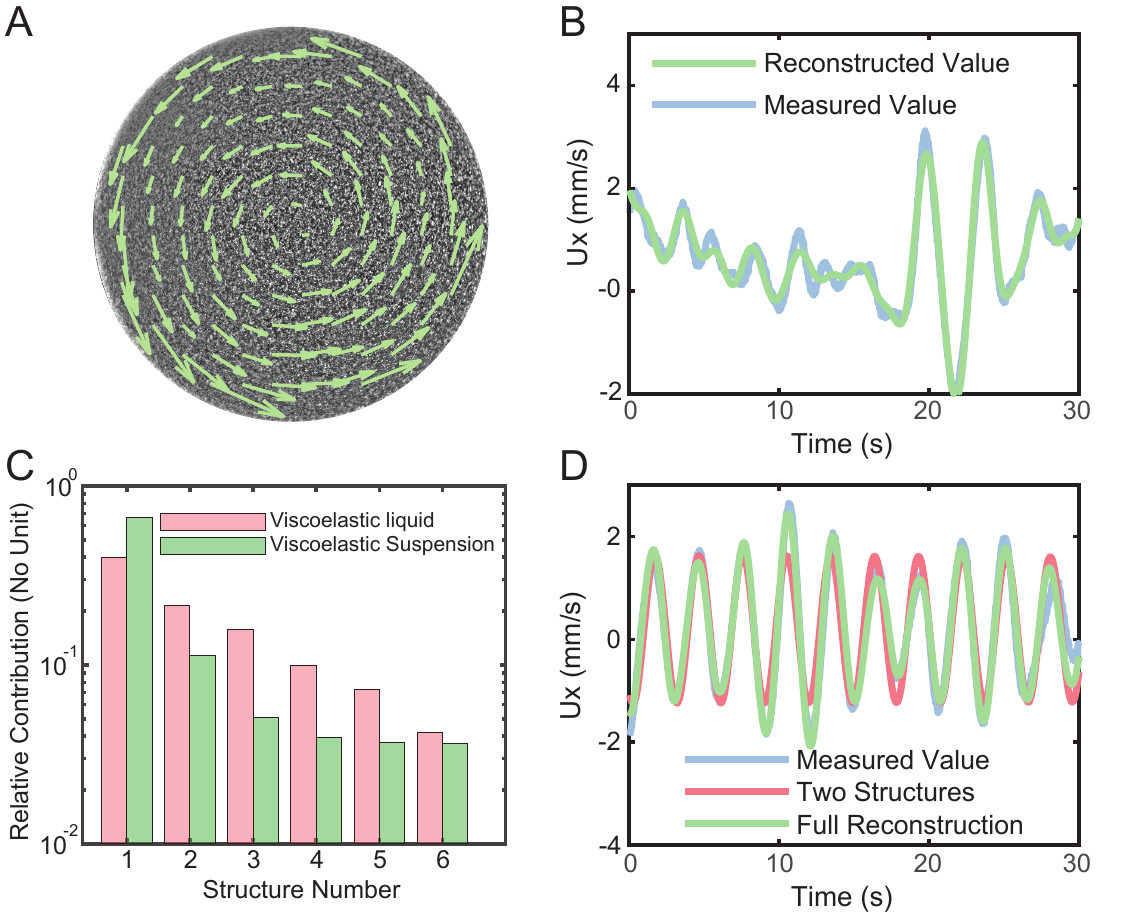}
\caption{Quantification of the viscoelastic secondary flow by PIV analysis and optDMD decomposition.
({\it A}) The fluorescent image of the polymer solution in the flow cell overlaid by the velocity field extracted from PIV analysis in a single frame. The particles move predominantly in the azimuthal direction.
({\it B}) The reconstruction of the velocity field of the viscoelastic liquid without particles from the optDMD method. The sky blue line is the measured velocity in the $x$ direction in time, averaged over the entire flow field. The jade green line is the reconstructed value  by the optDMD method based on 11 structures.
({\it C}) Comparison of the contribution of the coherent structures from the optDMD method, with and without the presence of the suspension. The addition of the suspension increases the contribution of the primary flow, which is Structure 1, suggesting the flow is less turbulent. Moreover, one single flow structure, structure 2, dominates over the others. 
({\it D}) The reconstruction of the velocity field of the viscoelastic suspension flow from the optDMD method.
The sky blue line is the measured velocity in the $x$ direction in time, averaged over the entire flow field. The pink line is the reconstructed value by the optDMD method with the primary and the first secondary flow structure. The jade green line is the reconstructed value based on all 11 structures.}

\label{fig:2} 
\end{figure*}
To better understand the impact of particles on the flow field, we employ particle image velocimetry (PIV) to directly measure the velocity field. For this experiment, we add a small amount (0.25 vol\%) of fluorescently labeled CA50 particles to the polymer solution. These particles are labeled with rhodamine B and are illuminated by a collimated LED with a wavelength of 505 nm and power of 220 mW (M505L4-C1 from Thorlabs). The velocity field is imaged with a camera (BFS-U3-32S4M-C from FLIR) at a frame rate of 200 fps. The velocity field exhibits a spiral-like pattern, as depicted in Fig.~\ref{fig:2}{\it A} and Movie S3.

To further analyze the spiral patterns observed in the velocity field, we apply optimized dynamic mode decomposition (optDMD) to identify a set of coherent structures in the flow field that each possess a single frequency and fixed growth rate. These structures characterize the fluid flow, with the first coherent structure corresponding to the primary flow and each additional structure contributing to the secondary flow~\cite{schmid2010dynamic, lee2021optimized}. By using optDMD, we are able to analyze more precisely the complex flow patterns present in the system.

The elastic instability in the polymer solution without particles is well described by the primary flow and ten secondary coherent flow structures. By summing all 11 coherent structures, we are able to adequately reproduce the measured velocity field, as demonstrated by the excellent agreement between the calculated and measured time dependence of the average horizontal velocity of the entire velocity field $U_x$, as shown in Fig.~\ref{fig:2}{\it B}. The first coherent structure, representing the primary flow, accounts for approximately 40\% of the total, while the next five structures, representing the secondary flow, each contribute at least a few percent, as illustrated in pale pink in Fig.~\ref{fig:2}{\it C}.

To investigate the effect of the particles on the flow field, we repeat the PIV and optDMD analysis using a mixture of 0.25 vol\% fluorescently labeled particles and 29.75 vol\% nonfluorescent particles. The resulting particle velocity field is again well described using 11 coherent structures, as shown by the sum of the velocity fields in Fig.~\ref{fig:2}{\it D}. However, we find that with the particles, the contribution of the first coherent structure becomes significantly larger, accounting for approximately 67\%. The second coherent structure becomes the predominant contributor to the secondary flow, with a contribution of 11\%, while the contribution of the remaining structures decreases significantly, as depicted in jade green in Fig.~\ref{fig:2}{\it C}.

The importance of the first and second coherent structures in describing the flow is further emphasized by their contribution to the time dependence of the average horizontal velocity of the entire velocity field $U_x$. When we decompose the flow into two structures, the resulting velocity field nearly matches the measured value. Adding the next nine structures results in only moderate improvement of the agreement, as shown in Fig.~\ref{fig:2}{\it D}. This highlights the significance of these two structures in describing the flow field in the presence of particles.

\begin{figure*}[htbp]
\centering
\includegraphics[width=7in]{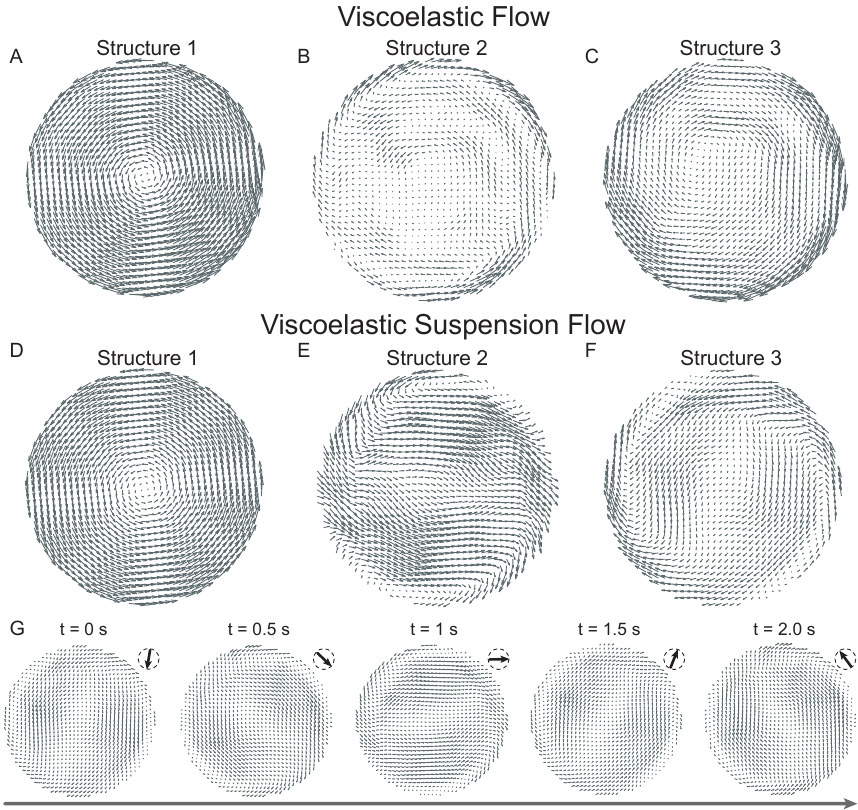}
\caption{Coherent structures of the flow based on DMD analysis. We plot the real part of the primary flow and the imaginary part of the secondary flow here.
({\it A}) The first coherent structure corresponding to the primary flow of the polymer solution. Most of the velocity vectors point in the azimuthal direction.
({\it B},{\it C}) The second and third coherent structures of the viscoelastic flow. The chaotic secondary flow consists of complex spiral structures, which propagate and intermingle over time{. The spirals mimic the patterns seen under rheoscopic measurements }.
({\it D}) The first coherent structure in the primary flow of the polymer solution with 30 vol\% suspension particles. Again, most of the velocity vectors point in the azimuthal direction.
({\it E}) The second coherent flow structure of the polymer solution with 30 vol\% suspension particles. The secondary flow consists of a rigid disk with a defect line in the middle and spiral patterns surrounding it.
({\it F}) The third coherent flow structure of the polymer solution with 30 vol\% suspension particles. The secondary flow is again a spiral whose velocity is smaller in the middle.
({\it G}) The evolution of ({\it E}) over time. The velocity field rotates over time like a rigid body.
}
\label{fig:3} 
\end{figure*}

For the polymer solution without particles, the velocity field of the primary flow is circular, with almost all velocity vectors pointing in an azimuthal direction, as depicted in the first coherent structure in Fig.~\ref{fig:3}{\it A}. The secondary flow has a spiral character, similar to the flow patterns observed with rheoscopic visualization, as shown in the higher-order structures in Fig.~\ref{fig:3}{\it B} and~\ref{fig:3}{\it C}. 

For the polymer solution with suspended particles, the velocity field of the primary flow is again circular, with almost all velocity vectors pointing in an azimuthal direction, as shown in the first coherent structure in Fig.~\ref{fig:3}{\it D}. Intriguingly, the velocity field of the secondary flow exhibits a very different character when particles are present. The velocity field is roughly constant near the center and rotates over time like a rigid body, surrounded by an outer layer of spirals, as shown in the second coherent structure in Fig.~\ref{fig:3}{\it E}, and Movie S4. However, the higher-order structures return to spiral patterns, as shown, for example, in the third coherent structure in Fig.~\ref{fig:3}{\it F}.

 Notably, in the presence of particles, the second coherent structure rotates at a frequency of 0.31 Hz, as shown in  Fig.~\ref{fig:3}{\it G}, which matches the frequency peak in the torque PSD (0.32 Hz). The swirling flow pattern observed for this sample with rheoscopy also has a frequency of 0.3 Hz. These findings suggest that the predominant contribution to the secondary flow in this system arises from the rotation of a nearly static structure in the center of the sample at a frequency of 0.3 Hz.

\section*{Rheomicroscopic visualization of the particle crystallization}
To further examine the apparent rotation of a nearly rigid body that contributes significantly to the secondary flow in the presence of particles, we observe the flow in the middle plane using a rheomicroscope (MCR 702 with the rheo-microscope accessory from Anton–Paar). This device consists of counter-rotating disks that maintain the middle plane stationary, allowing it to be visualized with a microscope and imaged with a high-speed camera (v7.3 from Phantom). By strongly shearing the sample with 30 vol\% particles up to a shear rate of $\dot{\gamma} \sim 10^2 ~\mathrm{s}^{-1}$ { in a smaller gap to minimize the scattering}, where the elastic instability is fully developed, the liquid is rapidly expelled, and we can only briefly visualize the particles. In doing so, we observe the formation of an ordered layer of particles in the stationary middle plane, as shown in SI Fig. \ref{fig:A3} and SI Video 6.

To study the ordering of particles in the midplane in more detail, we utilize the large-amplitude oscillatory shear (LAOS) protocol, which imposes a sufficiently large shear to induce structure formation, but, because it is oscillatory, does not expel the sample from the geometry ~\cite{hyun2011review}. We use an angular velocity $\omega = 2\pi~\si{rad/s}$ and a shear amplitude of $\gamma_0=3000\%$, resulting in an average applied shear rate of $\dot{\gamma}=2\omega\gamma_0/\pi=120~\mathrm{s}^{-1}$. The oscillatory motion also leads to the ordering of particles in the midplane, similar to that observed in steady shear. In this case, the ordering begins almost immediately, and after 60 seconds, a steady state is reached, where the torque applied to drive each oscillatory cycle becomes unchanged from cycle to cycle, suggesting that the particle structure has reached a steady state. In this steady state, the particles form an ordered structure that is retained through each cycle of LAOS, with the direction of particle motion reversing as the flow reverses each cycle. This allows us to visualize the details of the structure that gives rise to the secondary flow pattern. By observing the same area of particles in the flow geometry that remain in the field of view at the end of each cycle, we can track the configuration of these particles over time and gain insight into the structure contributing to the secondary flow pattern.

\begin{figure}[h!]
\centering
\includegraphics[width=4.5 in]{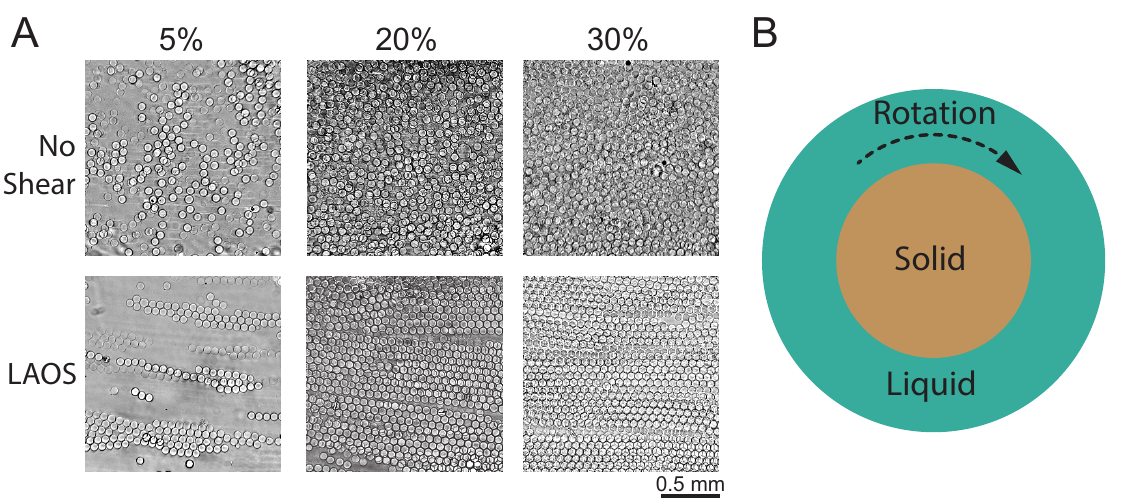}
\caption {Crystallization of the suspended particles under shear. ({\it A}) Rheo-microscopic visualization of the particles assembling under shear. We repeat the experiments with 5 vol\%, 20 vol\%, and 30 vol\% viscoelastic suspensions, shown as different columns. The first row is the control experiment without shear. Particles remain suspended. The second row is the same system as the first row, but the suspension is sheared under $1$ Hz oscillations with an amplitude $\gamma_0 = 30$. In the lower left, 5 vol\% particles chain up under shear, but there is space between the chains, which can still be considered suspended. In the lower middle, particles crystallize into 2D close-packed structures under shear. Those structures further overlap with each other, lock with their neighbors,  and move collectively in each LAOS cycle. In the lower right, a 30 vol\% suspension is sheared under the same condition. The suspension behaves similarly to the 20 vol\%  case. Particles crystallize into rafts. ({\it B}) Qualitative explanation of viscoelastic suspension under strong shearing. The suspended particles phase separate and crystallize into a solid core, which then rotates over time.}
\label{fig:4} 
\end{figure}

Using LAOS, we observe the assembly of particles even at a volume fraction as low as  $\phi$ = 5 vol\%. At this concentration, local structures consist of short chains and small rafts of these chains are formed. Remarkably, the particles become concentrated in the midplane of the sample. These structures further align with the flow and move in the flow direction. There is sufficient distance between the chain-like assemblies to allow relative motion between neighboring structures without collision. Therefore, although the particles assemble under shear, the resultant structures remain dispersed, occupying only about 20\% of the area, as shown in Fig.~\ref{fig:4}{\it A}.

Surprisingly, when the volume fraction of particles is increased to  $\phi$ = 20 vol\%, the attractive interactions among particles lead to phase separation. The particles crystallize into two-dimensional (2D) closely packed rafts instead of chains. These rafts migrate like a rigid body during each LAOS cycle. Due to the high concentration, the rafts overlap and become locked together, forming even larger structures, as seen in the lower middle of Fig.~\ref{fig:4}{\it A}. Increasing the volume fraction to  $\phi$ = 30 vol\% results in even larger rafts, but with qualitatively similar structures, as shown in the lower right of Fig.~\ref{fig:4}{\it A}. This crystallization is \add{anomalous}: in a monodisperse, hard-sphere suspension of Brownian particles, crystallization first occurs at  $\phi \approx 54.5\%$ ~\cite{hoover1968melting}, which is much higher than the 20\% particle volume fraction where we first observe rafts.

The formation of crystalline layers is due entirely to the interactions induced by the viscoelastic fluid; no such effects are observed for particles suspended in a Newtonian fluid. { Moreover, the particles are large enough that Brownian motion is insignificant; furthermore, there is no effect of a depletion interaction for these particles and polymer.} The elasticity of the fluid drives each particle to {directionally} migrate to the middle plane and further induces an attractive interaction between the particles that causes chain formation along the direction of flow. At low volume fractions, these interactions result in chains of particles being separated from one another, as seen in the lower left of Fig.~\ref{fig:4}{\it A}. As the volume fraction increases, the density of the chains also increases, eventually leading to the formation of crystals.

The formation of a crystal of particles is the cause of the behavior of the secondary flow reflected in the second coherent structure {and the change in the torque}. The crystal structure persists in the flow but does not fully align with the streamlines. Instead, a thin crystal plug forms in the center of the flow cell, resisting the viscoelastic shear flow and preventing the development of secondary flow in the middle, suppressing the elastic instability. Furthermore, the torque from the upper rotating plate drives the particle crystal to rotate at a constant rate, leading to the single frequency fluctuations seen in the torque measurements and in optDMD, as shown in Fig.~\ref{fig:4}{\it B}.

\section*{Polydisperse particles erase the regulation}

\begin{figure}[h]
\centering
\includegraphics[width=8.7 cm]{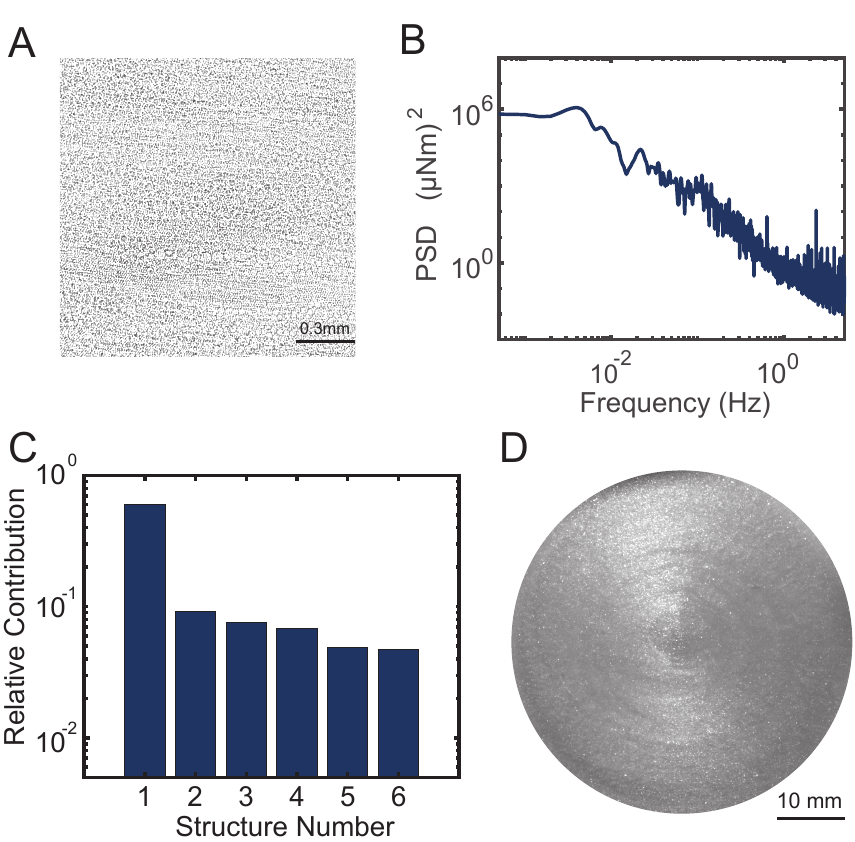}
\caption{Secondary flow of 30 vol\% polydisperse viscoelastic suspensions. ({\it A}) 30 vol\% polydisperse particle suspension in the LAOS test under the microscope. The in-plane volume fraction is high, suggesting the particles have attractive interactions and migrate toward the middle plane. However, there is no crystallization due to the polydispersity. ({\it B}) The PSD of the torque reading as a function of frequency. No sharp peak can be observed. ({\it C}) The optDMD analysis. In line with the rheoscopic visualization, in the polydisperse  suspension, no predominant mode can be seen. ({\it D}) The rheoscopic visualization of the sample under shear. Multiple spirals coexist, and no predominant motion is observed.}
\label{fig:5} 
\end{figure}

One of the most striking features of the structure is its crystalline order, despite the relatively low total volume fraction of particles. The particles are close packed, suggesting that the order arises due to the monodispersity of the particles. To test this hypothesis, we prepare a polydisperse suspension consisting of a mixture of 10 vol\% 9.2 micron-diameter, 10 vol\% 30 micron-diameter, and 10 vol\% 51 micron-diameter particles. When we perform LAOS measurements under the same conditions, we observe a concentrated layer of particles at the midplane, but with no crystalline order because of the polydispersity, as shown in Fig.~\ref{fig:5}{\it A}. ~\cite{pusey1987effect}. 

In addition, there is no single frequency peak in the power spectrum, as shown in Fig.~\ref{fig:5}{\it B}. The optDMD analysis also shows that the only dominant coherent structure is the first, while the subsequent five coherent structures each have similar amplitudes, as shown in Fig.~\ref{fig:5}{\it C}. The rheoscopic visualization reveals multiple indistinct spiral patterns, but no predominant pattern can be identified, as shown in Fig.~\ref{fig:5}{\it D}. This indicates that the secondary flow is suppressed by the suspension, while the swirling motion of the secondary flow spreads throughout the entire flow field. These observations confirm the critical role of crystalline order in controlling the secondary flow.

\section*{Conclusion}
This study has shown that the presence of particles in a viscoelastic polymer solution significantly alters the flow at high shear rates, leading to the formation of a crystalline layer of particles. This crystalline layer resists the viscoelastic shear flow and inhibits the development of the elastic instability, thereby regulating the secondary flow. Furthermore, the rigid body rotation of this crystal is responsible for the dominant frequency in the flow structure. These results have important implications for industrial settings where ordering is likely to modify to the processing of particle suspensions in viscoelastic fluids.  It would be very interesting to examine whether similar crystalline ordering occurs in other geometries.

\section*{Acknowledgement}
We are grateful for insightful discussions with Dr. Andreas Bausch, Dr. Jörg G Werner, Dr. Shima Parsa. We also acknowledge the support of the NSF through the Harvard MRSEC (DMR20-11754).
\bibliography{References.bib}
\subfile{Appendix}

\end{document}

%% file: Appendix.tex
\newpage
\section*{Supplementary Material}
\label{AppendixA}
\section*{Details of sample preparation and experimental setup}

We prepare a polymer solution by mixing 25.814g of DI water with 141.765g of 2-2 thiodiethanol (TDE) and storing it in a -20°C freezer to prevent clumping of the undissolved polymers. We then add 0.3850g of PEO with a molecular weight of $8\times10^6$ Da to the mixture and heat it in a 65°C oven for 4 hours to dissolve the polymers. We then homogenize the mixture by gently shaking it with a mechanical shaker for 30 minutes. The resulting polymer solution has a density of 1.18 g/ml at 20°C. To form the suspension, we add 68.2g of surfactant-free PMMA particles (CA 50 from MICROBEADS) with a density of 1.19 g/ml to the polymer solution, resulting in a final volume fraction of 30\%. The particle size standard deviation is less than 1 \textmu m, according to the manufacturer. For experiments without particles, we use a polymer solution of equal weight to replace the particles.

For the Particle Imaging Velocimetry (PIV), 0.83 wt\% of the particles (0.25 wt\% of the total suspension) are replaced with fluorescent particles.  To ensure the fluorescent particles have the same physical property as other nonfluorescent particles and strong fluorescence to be detected with a commercial camera, we exploit the polymer swelling to introduce Rhodamine B dye to the CA50 particles. The particles are firstly immersed in a 10 g/L Rhodamine B isopropanol solution and heated in  a 65 °C oven for 24 hours to swell the particles and let the dye molecules diffuse into the particles. We then rinse the particles with room temperature deionized water five times to remove the residual dye in the solution and let the particles deswell. The dye is then trapped in the particles and emits strong fluorescent light. 

To carry out  PIV experiments, we outfit the MCR 501 rheometer with customized optics for flow visualization. Our design is inspired by previous rheometric flow visualization studies ~\cite{byars1994spiral,pommella2019coupling}. We replace the lower plate of the rheometer with optical-clear laser-cut 6.35 mm thick plexiglass (8560K354 from Mcmaster-Carr). The 50 mm diameter upper plate (\# 12081 from Anton Paar) is attached with an OD4 absorptive Neutral Density filter (\# 36-276 from Edmund Optics) to eliminate undesired background light. The top and bottom plates were carefully aligned, with a height difference of less than 0.05mm across the plates, as measured with a feeler gauge. We add a 75 mm diameter cup to the lower plate. 200 ml sample will fill the cup up to around 45mm. In the torsional shear flow, the velocity gradient is inversely proportional to the gap size, so the velocity gradient between the upper and lower plate with a 2 mm gap is at least ten times larger than the gradient between the upper plate and the top free surface, whose gap size is around 40 mm. It ensures the torque reading is mainly contributed by flow beneath the upper plate, and the flow above the upper plate remains laminar.

\section*{Optical Imaging}

We use a table lamp (NÄVLINGE from IKEA) as the light source for rheoscopic visualization. For fluorescent imaging, we choose a 505 nm wavelength, 220 mW  collimated LED (M505L4-C1 from Thorlabs) as the light source. The green light beam is further expanded by a light diffuser (ED1-C20-MD from Thorlabs) to a 50 mm diameter. 

We mount an 1/1.8 inch C-mount camera (BFS-U3-32S4M-C from FLIR) with 35mm focal length lens with f-number $f_\mathrm{num} = 1.4$ (MVL35M1 from Thorlabs) for imaging. A 550 nm long-pass filter (FEL0550 from Thorlabs) is attached to the lens. It can capture the flow field with approximately 700 px by 700 px resolution in 200 fps under pixel binning mode. The actual sensor size of the camera is a square with a side length of about $l_i =5$ mm, and the pixel size is 7 {\textmu}m. The magnification ratio $M=l_i/{2R}=0.1$. If we consider the pixel size as the diameter of the circle of confusion $d_c=7$  \textmu m, the depth of field (DoF) can be estimated as 
$$DoF \approx \frac{2f_\mathrm{num} d_c}{M^2}=\frac{2\times1.4\times0.007}{0.1^2}=2.2~\mathrm{mm},$$

which is comparable with the gap height $H = 2$ mm. Therefore, all the particles are in focus.

\section*{Details about PIV and DMD decomposition}

\subsection{Details about velocity measurement}

For the flow field measurement, we record particle images at 200 fps after the torque reading from the rheometer reaches statistically stationary, which is after about 60 seconds. {Because of the directional migration of particles towards the middle plane in viscoelastic flow, the velocity is measured in the middle plane of the flow cell.} The raw image is processed with the Matlab PIVLab plugin. We apply the CLAHE filter with a 64 px by 64 px window size and a high-pass filter with a 15 px kernel size as a pre-processing step to enhance the contrast. We obtain the velocity vectors by using iterative 2D cross-correlations with multiple sizes of interrogation windows. We use 50\% overlapped 64 × 64 pixels for the coarse grid and 50\% overlapped 32 × 32 pixels for the fine grid. We also check the cross-correlation map, where we find the signal-to-noise (SNR) ratio is about 2.
\subsection{Error Estimation}

\begin{figure*}
\centering
\includegraphics[width=6 in]{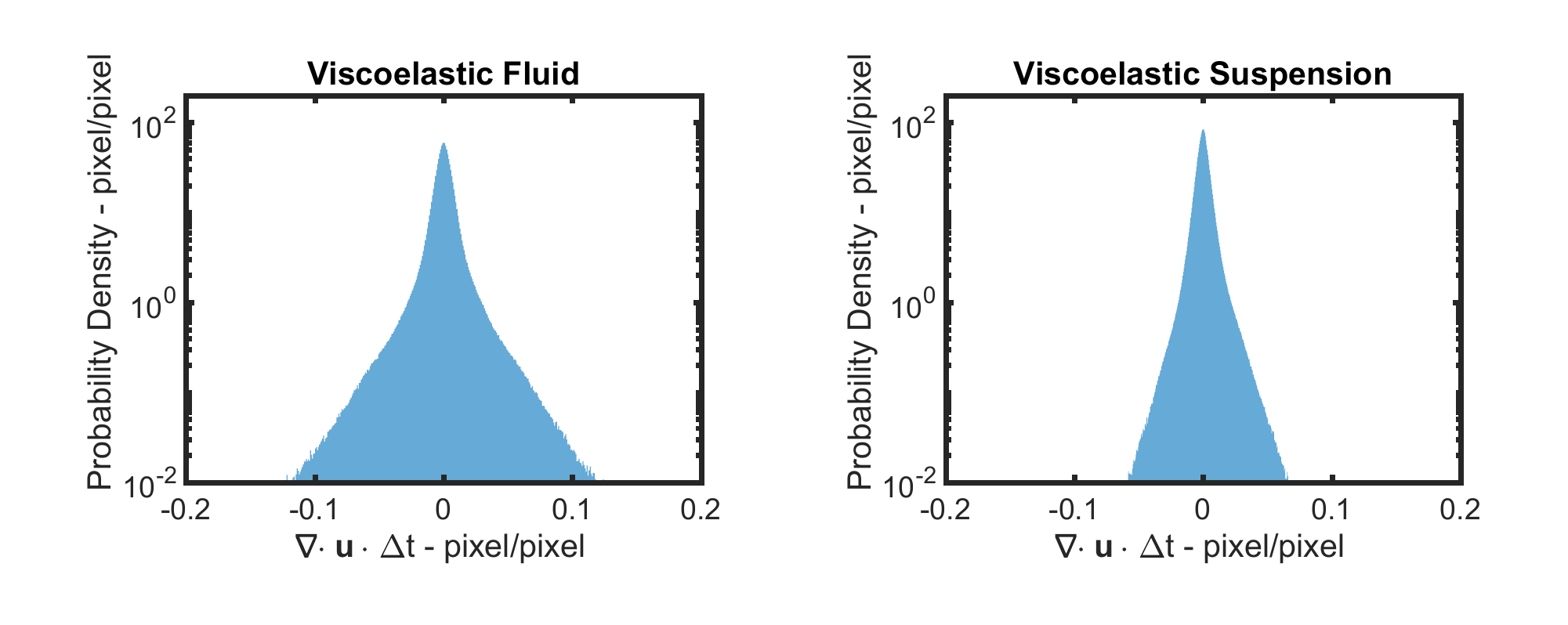}
\caption{The distribution of the residual in two velocity measurements. The standard deviation of the left side is 0.0134 pixel/pixel, and the right side is 0.0070 pixel/pixel. Both values correspond to less than a 1-pixel measurement error, suggesting our measurement is reasonably accurate.}
\label{fig:A16}
\end{figure*}

{We utilize the property of incompressible flow to estimate the error in our velocity measurements\cite{adrian2011particle,kim2011full}. Assuming the measured flow field at the midplane, when reaching statistical equilibrium, is a 2D flow; the net flow in the z (height) direction is zero. Consequently, $\nabla \cdot \mathbf{u} = 0$, where $\mathbf{u}=(u_x,u_y)$, where $u_x$ is the x directional flow velocity and $u_y$ is the y directional flow velocity. Guided by this principle, we can provide an order-of-magnitude estimate of the velocity measurement error as follows:
$$Residual~=~\Delta t \nabla \cdot \mathbf{u} $$;
the time interval $\Delta t = 0.005~\si{s}$ for 200 fps measurement. We calculate the residual distribution per point per frame for the polymer solution flow and the viscoelastic suspension flow, as seen in Fig. \ref{fig:A16}. The relative error can be estimated from the standard deviation of the residual $\sigma_r=\mathrm{std}(Residual)$, where $\sigma_r = 1.3\%$ for the polymer solution and $\sigma_r = 0.7\%$ for the viscoelastic suspension. 
}

\subsection{optDMD decomposition}

We then treat the time-dependent velocity field as a time series and process it with the optDMD method. If a coherent mode is not stationary, like those belonging to the secondary flow, then the eigenvector has an imaginary part. And the complex conjugate of the mode is also a coherent mode, resulting in two coherent modes per coherent secondary flow structure. This fact drives us to choose odd-number modes to decompose the flow. And combine the physically identical complex conjugated modes afterward. The relative error decreases with an increase in the number of modes used in the decomposition until the number of modes $n_m = 21$, or 1 primary flow plus 10 secondary flow structures. Adding more modes does not provide significant additional benefits. And We cut off at $n_m = 21$.

\begin{figure*}
\centering
\includegraphics[width=3 in]{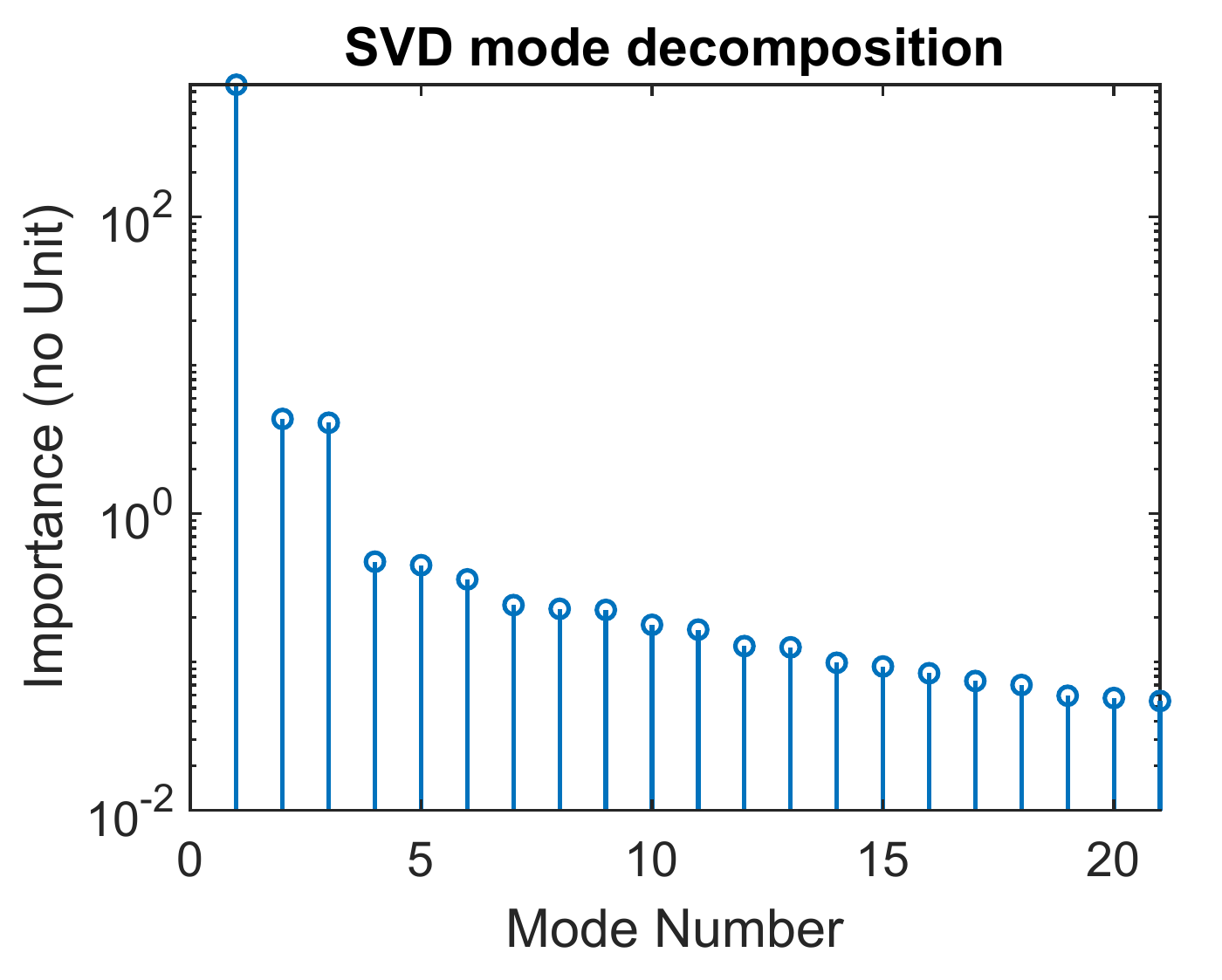}
\caption{The squared singular value of the first 21 modes in the SVD decomposition of the suspension flow. Mode 2 \& 3 dominate over other modes in the secondary flow, in line with our finding with DMD decomposition}
\label{fig:A12}
\end{figure*}

The argument that the viscoelastic suspension flow has a predominant coherent motion can be cross-validated by other postprocessing methods, like the Singular Value Decomposition (SVD). The SVD is a deterministic method and has no fitting parameter. If we sort the SVD  modes of the viscoelastic suspension flow by their importance, defined by the squared singular value, we can also find a pair of modes that dominate over the others, as seen in Fig. \ref{fig:A12}. And its  shape or eigenvector is the same as the solution of the optDMD method. 
\section*{Details of Rheo-microscopy}
To visualize the sample in the flow geometry {\it in situ}, we use the MCR 702 rheometer from Anton Paar with a counter-rotating concentric plate-plate geometry. The transparent upper and lower plates have a radius of 21.5 mm. Using the high-speed camera set at 100 fps, we visualize the local flow approximately 14 mm from the center, roughly 2/3 of the radius. The local shear rate, where the flow is observed, is nearly equivalent to the nominal shear rate in the parallel plate geometry, as determined by the software input. To minimize scattering from the dense suspension, we set the gap to 0.2 mm. The sample is illuminated with a white light lamp through a pinhole to enhance the contrast. We use a high-speed camera (Phantom V 7.3) with a pixel size of 22 \textmu m and a resolution of 800 px by 600 px { and 100 FPS. We find 100 FPS  delivers shape image of the sample at the end of each cycle} We mount the camera with a 5X objective (Mitutoyo), leading to a rectangle field of view of 3.43 mm by 2.57 mm. In postprocessing, we normalize the illumination and enhance the contrast to better quantify the particles. The surface fraction of the particles is determined through manual counting.

\section*{Onset of the elastic instability}
\begin{figure*}[h]
\centering
\includegraphics[width=0.5\textwidth]{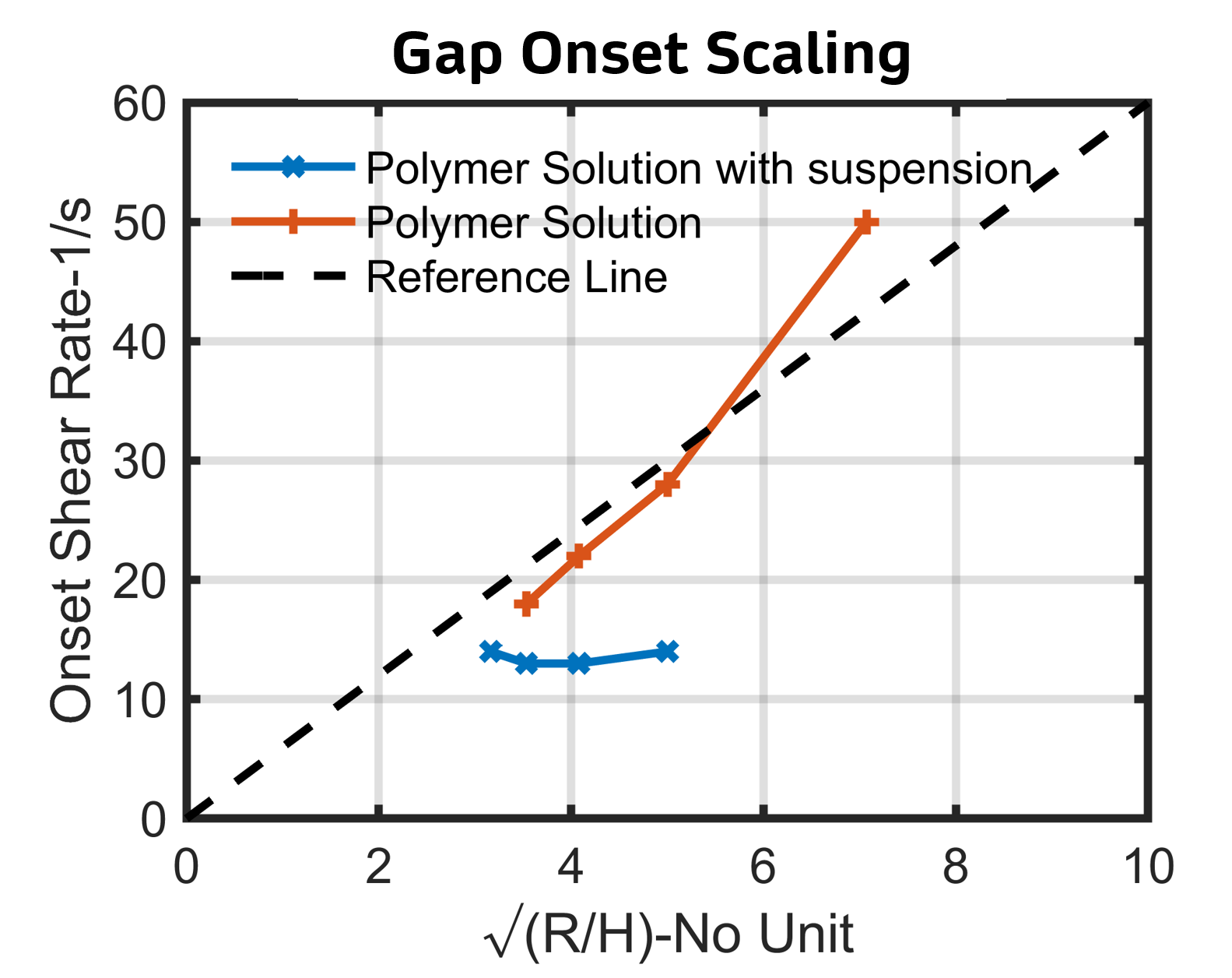}
\caption[Influence of the particle suspension on the onset of Elastic instability]{\label{fig:a1} The onset of Elastic instability. The square root of the radius $R$ to height $H$ ratio is the x-axis, and the instability onset shear rate is the y-axis. The black reference line is the theoretic prediction from McKinley, Oztekin 1996~\cite{mckinley1996rheological}. With the suspension, the critical shear rate for elastic instability is independent of the geometry.} 
\label{fig:A1}

\end{figure*}

The stability of the flow in a viscoelastic fluid is dependent on the ratio of elastic stress to viscous stress, which is governed by the shear rate and the relaxation time of the material. Additionally, the stability also depends on  the geometry of the flow cell. Previous studies conclude that the onset of the elastic instability in torsional shear flow can be described by nondimensional criteria $M_0$~\cite{mckinley1996rheological}.
$$\tau \dot{\gamma}_\mathrm{crit} \sqrt{H/R} > M_0 ,$$

where $\tau$ is the relaxation time of the material, $R/H$ is the radius-to-height ratio of the flow cell,  $\dot{\gamma}_\mathrm{crit}$ is the critical shear rate for flow instability and $M_0$ is a constant. We compare our results with the prediction. Our findings show that, for measurements with varying gap heights, $\dot{\gamma}_\mathrm{crit}$, where shear thickening first occur in viscosity measurement, scales with $\sqrt{R/H}$ with a gentle upward deviation, or the product $\dot{\gamma}_\mathrm{crit}\sqrt{H/R}$ remains roughly constant at the onset of elastic instability, as depicted in Fig.~\ref{fig:A1}. This reconfirms the validity of the $M_0$ criteria. The upward deviation observed may be attributed to the decrease in the relaxation time of the polymer solution as the shear rate increases.

Interestingly, when we add particles to the polymer solution. The critical shear rate $\dot{\gamma}_\mathrm{crit}$ where the shear thickening first occurs becomes independent of the flow geometry, as shown in Fig.~\ref{fig:A1}. Such change is because of the nature of the suspension flow, the suspension has two different inherent length scales: the length scale of each particle and the length scale of the entire flow geometry. The elastic instability can develop in both length scales, as the streamline is curved in both cases. 

Our data suggest that the flow change first occurs at each particle scale, generating additional dissipation and causing shear thickening. Similar observations have been reported in previous studies~\cite{yang_mechanism_2018}. Further increasing the shear rate, the instability eventually develops in the flow geometry scale. 

{ We can also rethink this question dimensionlessly. An important dimensionless number in viscoelastic flow is the Weissenberg number: $Wi = \dot{\gamma}\tau$, which characterizes the elastic stress to viscous stress ratio locally at each point. The $M_0$ criterion can be rewritten as $M_0=Wi\sqrt{H/R}$. Notice $Wi$ is scale-independent, and the local configuration of the flow channels between particles can be versatile. Thus, the instability can initialize at the particle scale under the most favorable condition, and $\sqrt{H/R}$ is irrelevant in this case. Thus, at the particle level, the $M_0$ criteria degrade to the $Wi$ criteria. The critical Weissenberg number in our case is $ Wi \approx 6$.}

{\section*{Deborah number and the scaling of the dominant frequency}}
{ The Weissenberg number, $Wi = \dot{\gamma}\tau$, is a dimensionless number that is defined based on the stress ratio as a local property. The Deborah number, $De=\Omega\tau$, is a second dimensionless number important for viscoelastic flow that is defined at the entire flow cell level. Here, $\Omega$ is the angular velocity of the upper plate. In the torsional shear flow between two parallel plates,$\dot{\gamma}=\frac{2R}{3H}\Omega \propto \Omega$, so $Wi \propto De$. However, we can decouple these two dimensionless numbers by alternating the height $H$ or radius $R$ of the flow cell.
} 

\begin{figure*}[h]
\centering
\includegraphics[width=0.75\textwidth]{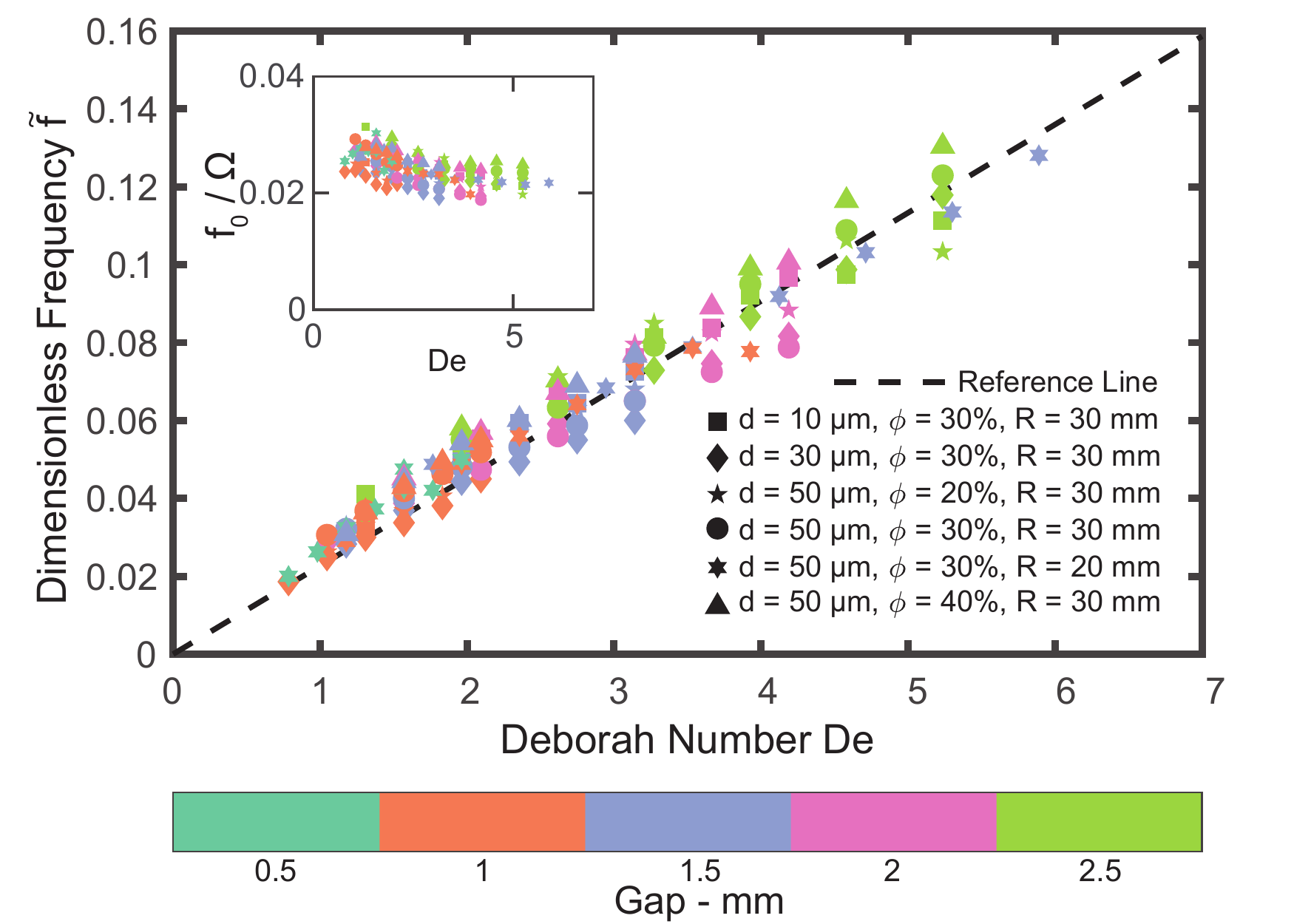}
\caption{Scaling of the dimensionless peak frequency $\tilde{f}=f_0\tau$ as a function of $De$, we approximate the relaxation time $\tau = 0.5~\si{s}$ based on the $N_1$ measurement stated later. The data is measured with a DHR-3 rheometer and 40 mm and 60 mm parallel plate geometry. Here we find the $\tilde{f}$ scales near linearly with $De$ with a gentle decline.} 
\label{fig:A14}

\end{figure*}

{We examine the scaling of the dominant frequency $f_0$ by repeating the experiment across a wide range of experimental conditions, including different shear rates, different particle sizes, different volume fractions, different geometry radii, and different gap heights. We find that the dimensionless frequency $\tilde{f}=f_0\tau$ is dictated by $De$ with a near-linear relationship, as depicted in Figure~\ref{fig:A14}. This dependence contrasts sharply with the onset of the shear thickening, which is dictated by $ Wi$. }

{The transition from $Wi$-dominant at the onset to $De$-dominant at the fully developed secondary flow manifests that the particles suspended in a viscoelastic fluid crystallize into an assembly under strong shearing. The scale of the assembly is comparable to the flow cell, and the single particles are depleted in the system, making the Weissenberg number less relevant when considering the fully developed secondary flow.}

\begin{figure*}[h]
\centering
\includegraphics[width=0.75\textwidth]{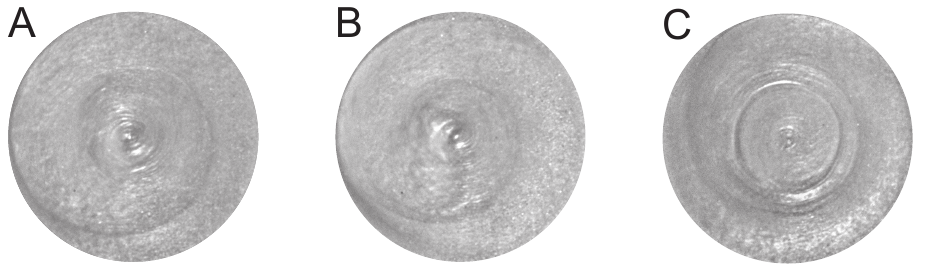}
\caption{Viscoelastic flow visualization under different conditions. Here 30 vol\% viscoelastic suspension is tested with the MCR 501 rheometer, similar to Fig 1.E. A)  Rheoscopic visualization of the secondary flow under condition $\dot{\gamma}=50~\mathrm{/s}$ and $H = 2~\mathrm{mm}$. B) Control experiment with $\dot{\gamma}=100~\mathrm{/s}$ and $H = 2~\mathrm{mm}$. C) Control experiment with $\dot{\gamma}=100~\mathrm{/s}$ and $H = 1~\mathrm{mm}$,} 
\label{fig:A15}

\end{figure*}

{\section*{Robustness of the measurement}}
{ In the working condition of most of the measurements, the secondary flow is fully developed, and we can observe similar flow behavior if we further increase the shear rate. The highest shear rate accessible with our current setup is 200/s, and we observe similar single-frequency torque oscillation following the same scaling rule in Fig.~\ref{fig:A14}. We also repeat the rheoscopic visualization under different conditions, where we find similar results, as depicted in Fig.~\ref{fig:A15}.}

\section*{Rheology of the material}
\subsection{Relaxation Time}

When the dispersed long-chain polymer deforms under shear, it takes time to recover, and this time scale is the physical origin of the relaxation time of the polymer solution. In practice, the polymer chain usually has different deformation modes, and each mode can have its own relaxation time, making it difficult to define a single material constant $\tau$ as the relaxation time for the polymer solution. Here we report two different methods to determine the relaxation time. 

In the first method, we measure the relaxation time of the polymer solution with the stress relaxation test. Here we apply a fixed amount of strain to the polymer solution and record the stress relaxation process with the Discovery HR-3 rheometer from the TA instruments. The stress relaxation of a linear viscoelastic fluid (Maxwell fluid) should be:
$$\sigma(t)=\sigma_0 \exp{(-\frac{t}{\tau})},$$
where $\sigma_0$ is the applied shear stress at $t = 0~\si{s}$, $\sigma(t)$ is the measured shear stress as a function of time $t$. Thus we can find $\tau$ by the following linear fitting:
$$\log{\sigma(t)}=\log{\sigma_0}-\frac{t}{\tau},$$
In the experiment, we find that the logarithmic stress does not linearly relax over time, but we can still obtain a single relaxation time by fitting the stress relaxation process from 0.1 s to 1.0 s,{ whose reciprocal spans from 1/s to 10/s, comparable to the shear rate range used in $N_1$ measurement}. The result is about $\tau =1.0 ~\si{s}$.

In the second method, we focus on the axial force $F_\mathrm{axial}$ measured by the DHR-3 rheometer with a 60 mm 2.0° cone-and-plate geometry. The measured axial force from the rheometer upper plate can be exploited to calculate the first normal stress difference

$$N_1=\frac{2F_\mathrm{axial}}{\pi R^2},$$

The first normal stress difference $N_1$ is again the manifestation of the elasticity of the dispersed long-chain polymer. And elastic instability is a normal stress effect. So the relaxation time  calculated based on $N_1$ is more relevant to elastic instability. For a viscoelastic material, the relaxation time can be calculated as follows:

$$\tau=\frac{N_1}{2\eta\dot{\gamma}^2},$$

Because of shear thinning and other effects, the measured viscosity $\eta$ and $N_1$ both depend on the shear rate $\dot{\gamma}$. So the relaxation time calculated based on this formula is also a function of the shear rate. And the data at $\dot{\gamma}=50~\si{s}^{-1}$ is inaccessible because the shear flow is unstable under such a high shear rate. Nevertheless, we can still obtain a relaxation time based on the rheometric flow at $\dot{\gamma}=10~\si{s}^{-1}$. The relaxation time $\tau = 0.64$ s for the polymer solution.

Upon the addition of the particles, the measured relaxation time from both methods decreases. The stress relax faster with the addition of the particles. The relaxation time with linear regression decreases from 1.0 second to 0.8 seconds. And the measured relaxation time based on the normal stress decreases from 0.64 to 0.49 sec at shear rate $\dot{\gamma}=10~\si{s}^{-1}$. See the figures for more details.
\begin{figure*}
\centering
\includegraphics[width=1\textwidth]{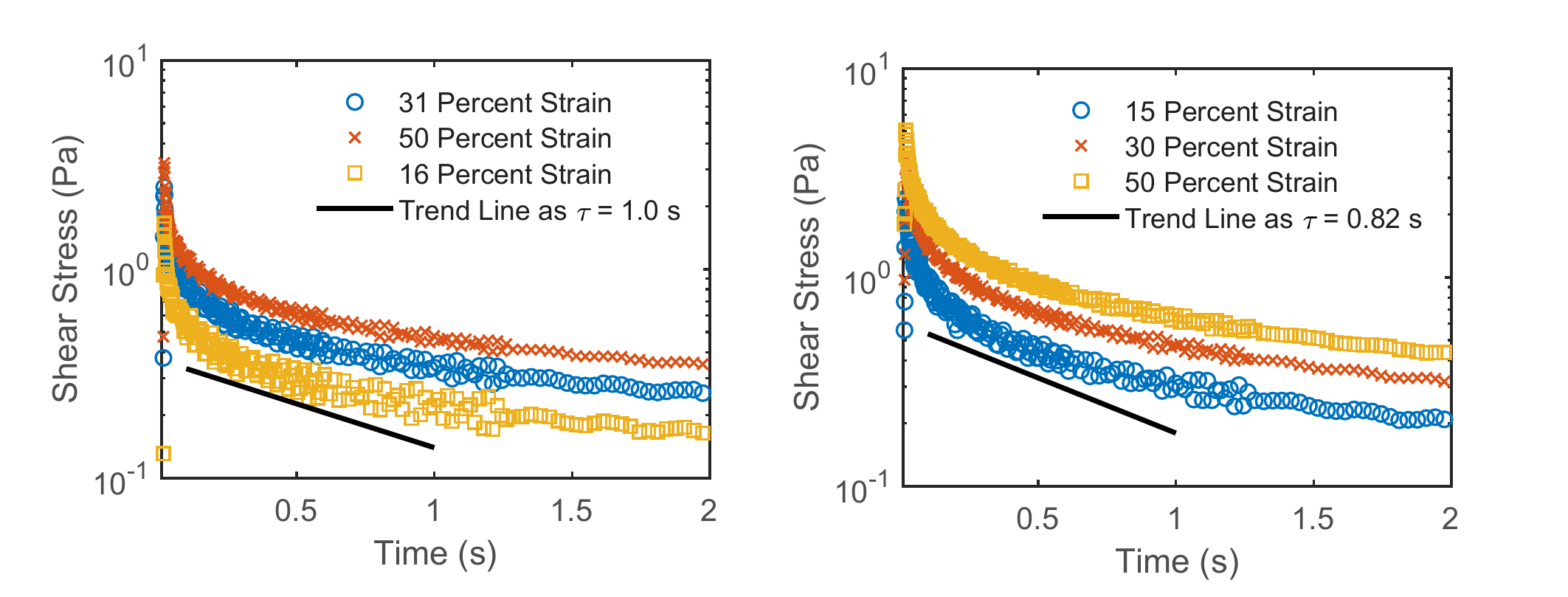}
\caption{ Stress relaxation test to determine the relaxation time of the system. The left side is the polymer solution that relaxes its stress over time under different shear strain loadings; the stress relaxation from 0.1 s to 1 s is linearly fitted to determine the relaxation time, highlighted as the black line here, { the stress relaxation over three different applied strain exhibit similar rate}. The fitted slope is 1.0 s for all three cases. The right side is the same experiment with a polymer solution containing 30 vol\% particles. The relaxation time decreases from 1.0 second to 0.8 second.   }
\label{fig:A9}
\end{figure*}

\begin{figure*}
\centering
\includegraphics[width=1\textwidth]{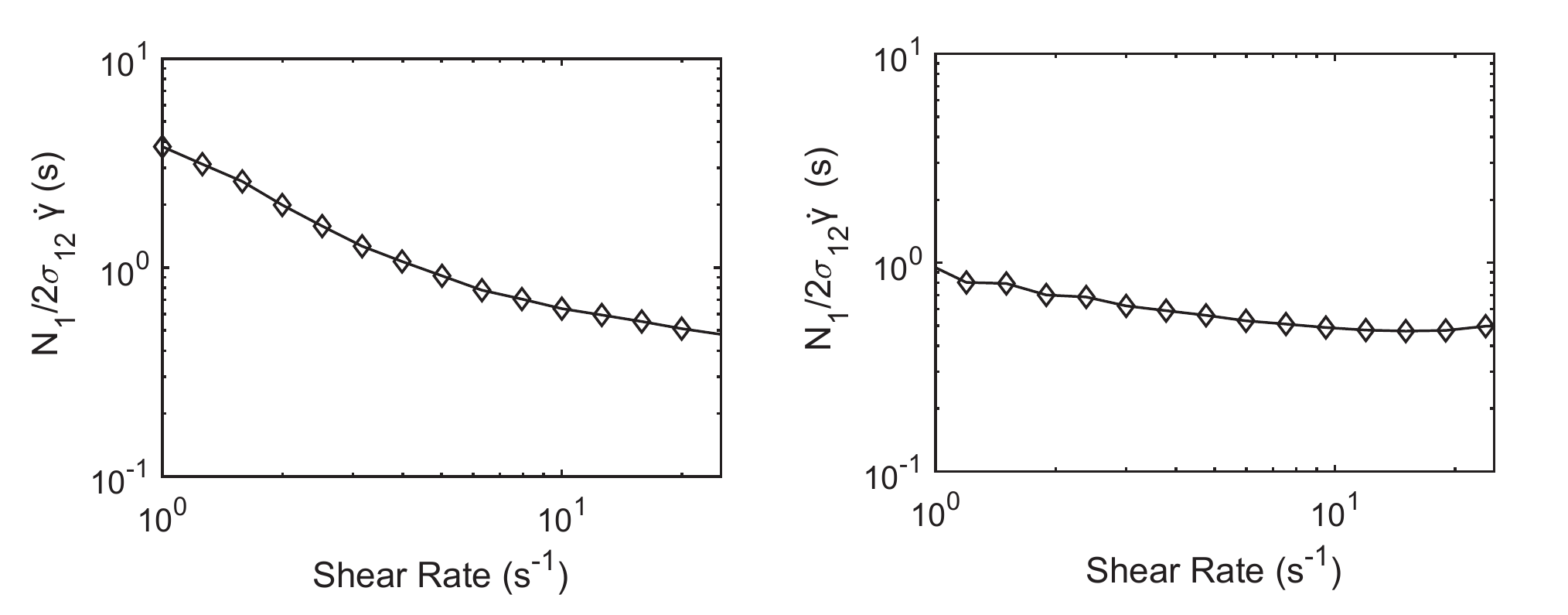}
\caption{ First normal stress difference $N_1$ divided by the shear stress $\sigma_{12}$ and the shear rate $\dot{\gamma}$ as a function of of $\dot{\gamma}$. The left side is the data from the polymer solution without particles, the right side is the data measured with a polymer solution and 30 vol\% 9 \textmu m diameter particles. The data is collected with a DHR-3 rheometer with a 60 mm 2.0° cone-and-plate geometry. The truncation gap size is 59 \textmu m, so we use 9 \textmu m diameter particles here. The polymer solution has a relaxation time $\tau=\frac{N_1}{2\eta\dot{\gamma}^2} = 0.64$ s at $\dot{\gamma} = 10/$s. And the polymer solution with 30 vol\% particles has a relaxation time $\tau=\frac{N_1}{2\eta\dot{\gamma}^2} = 0.49$ s  at $\dot{\gamma} = 10/$s}
\label{fig:A8}
\end{figure*}

\subsection{Viscosity measurement}

We measure the viscosity as a function of the shear rate for both polymer solution and polymer solution with 30 vol\% suspensions. For the polymer solution, we use a 60 mm cone-and-plate geometry on a DHR-3 rheometer to measure the viscosity and the normal stress as a function of the shear rate. We find the viscosity exhibit shear thinning  above $\dot{\gamma}\sim 0.1~\si{s}^{-1}$. The viscosity is about $1.1$ Pa.s when the shear rate $\dot{\gamma} = 1~\si{s}^{-1}$.The measured viscosity keeps decreasing until the onset of the elastic instability.

We also measured the viscosity of the polymer solution with 30 vol\% 9 \textmu m diameter particles with the same setup. The addition of the particles increases the measured viscosity from $\eta = 1.1$ Pa.s to $\eta = 2.8$ Pa.s under $\dot{\gamma} = 1~\si{s}^{-1}$.

\begin{figure*}
\centering
\includegraphics[width=1\textwidth]{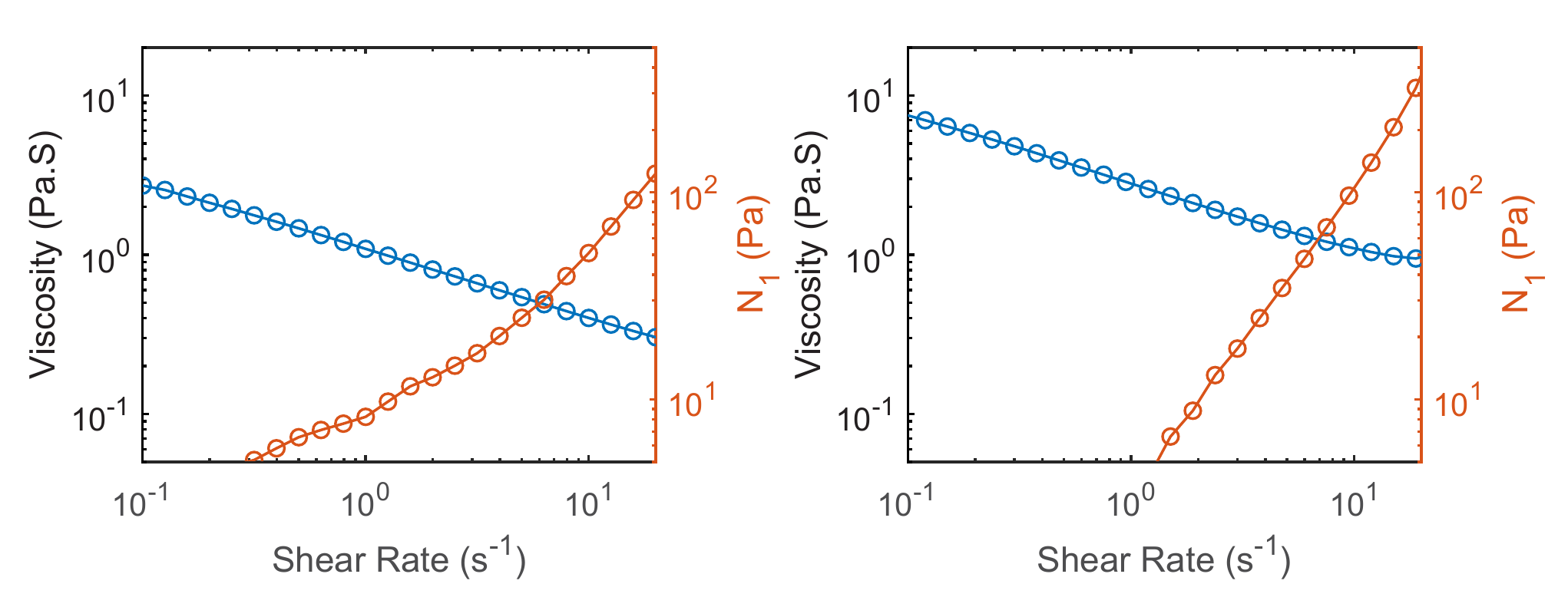}
\caption{Viscosity measurement of the polymer solution and polymer solution with 30 vol\% 9 \textmu m diameter particles, measured with a 2° 60 mm cone-and-plate geometry on the DHR-3 rheometer from TA instrument. The left side is the polymer solution with no particles; The right side is the polymer solution with 30 vol\% particles}
\label{fig:A8}
\end{figure*}

\begin{figure*}
\centering
\includegraphics[width=1\textwidth]{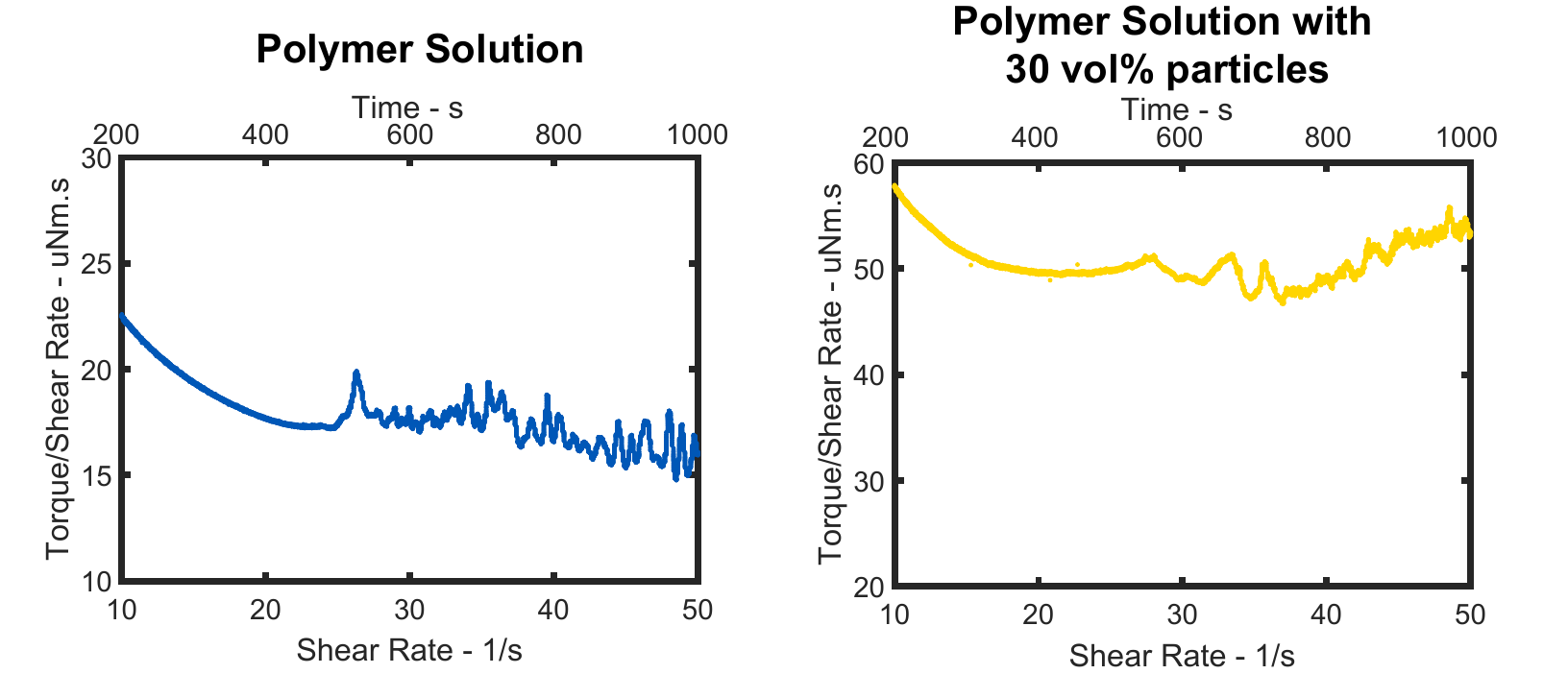}
\caption{Shear rate ramp of a viscoelastic fluid and the same fluid with 30 vol\% particles from 10/s to 50/s over 800 s in an immersion flow cell. The setup is similar to that of Figure 1. In the viscoelastic fluid (left), the torque begins to fluctuate immediately after shear thickening occurs, suggesting that the instability develops at the flow cell level. In the viscoelastic suspension (right), shear thickening happens before the torque fluctuation begins, suggesting that the flow instability develops from the particle level and eventually develops at the flow cell level.}
\label{fig:A8}
\end{figure*}

\subsection{{The size-dependence} and surface effect of the particles}

\begin{figure*}
\centering
\includegraphics[width=1\textwidth]{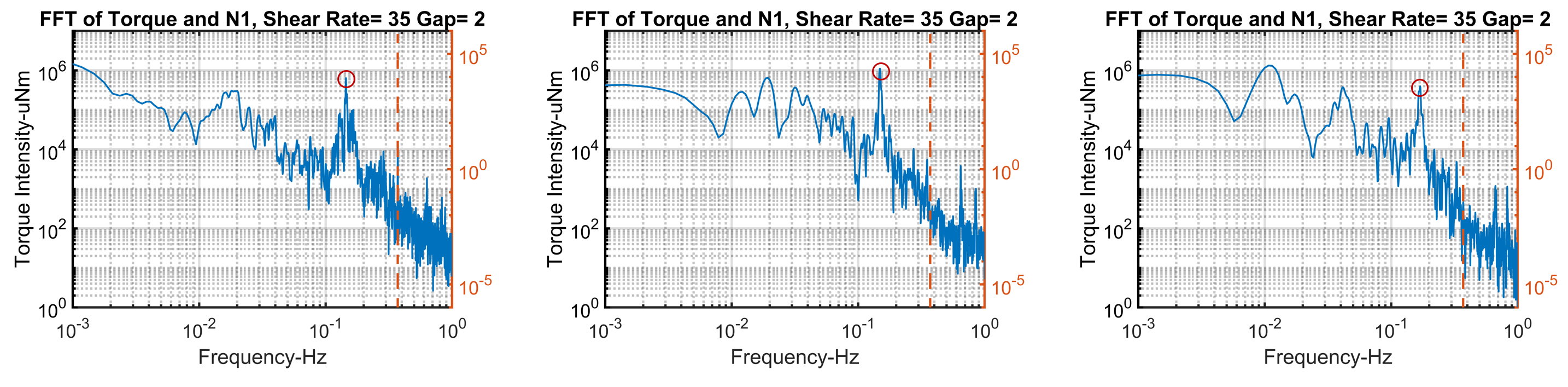}
\caption[The influence of the particle size on the secondary flow,]{\label{fig:a4} The influence of the particle size on the secondary flow, the power spectrum is the same with different particle sizes (51 {\textmu}m on the left, 30 {\textmu}m in the middle and 9 {\textmu}m on the right). The shear rate here is 35 /s and the gap size is 2 mm with a 60 mm plate-plate geometry, measured with a DHR-3 rheometer} \label{fig:A11}
\end{figure*}

The diameter of the particle we choose in this study ranges from 10 ~\textmu m to 51 ~\textmu m. We focus on the hydrodynamic interactions between the suspended particles. To verify that they dominate over other surface interactions, {and the size of the particle is less relevant, }we test particle suspension with the same 30 vol\% but different particle sizes, resulting in different surface-to-volume ratios. In the torque measurement, we find very similar power spectra, suggesting other surface interactions are less relevant in this study, as shown in Fig. \ref{fig:A11}. 
\section*{Additional explanation towards particle assembly in viscoelastic flow}

The assembly of particles into 2D aggregates is observed in both steady shearing and oscillatory shearing, as seen in SI Video 6 and SI Video 9. This assembly can be contextualized by previous studies in the field.\cite{becker1996sedimentation,d2015particle,feng1996motion,murch_collective_2020,won2004alignment,scirocco_effect_2004,xie2016flow,pasquino2010directed}

Previous studies have shown that spheres in viscoelastic fluids experience a force perpendicular to a wall, which causes them to move away from the wall. Single or multiple particles can migrate away from the wall in this way. In a torsional flow cell, the presence of two walls causes the particles to migrate toward the midplane.

Furthermore, viscoelastic fluid flow generates an attractive interaction between particles. The major factor is the normal stress perpendicular to the streamlines exerted by long-chain polymers. As two particles approach each other and their local streamlines overlap, this normal stress leads to a net force on each particle, pushing the particles toward each other. 

Considering these two factors, we argue that the viscoelasticity of the fluid drives particle assembly under shear. Evidently, no directional migration nor attractive interactions are observed in Newtonian suspension under similar conditions.
\section*{SI Videos}

\textbf{SI Video 1 (separate file)} - Rheoscopic visualization of Polymer Solution Flow: This video demonstrates the flow of a polymer solution visualized using rheoscopic techniques. The video is in real-time, and the diameter of the circle is 50 mm.

\textbf{SI Video 2 (separate file)} - Rheoscopic visualization of Viscoelastic Suspension Flow: This video demonstrates the flow of a viscoelastic suspension visualized using rheoscopic visualization methods. The video is in real time, and the diameter of the circle is 50 mm.

\textbf{SI Video 3 (separate file)} - Polymer Solution Velocity Field: This video displays the velocity field of a polymer solution  under Shear Rate = 50/s. The velocity is calculated per 32 px by 32 px square with a 16 px step size.

\textbf{SI Video 4 (separate file)} - Viscoelastic Suspension Velocity Field: This video displays the velocity field of 30 vol\% viscoelastic suspensions under Shear Rate = 50/s. The velocity is calculated per 32 px by 32 px square with a 16 px step size.

\textbf{SI Video 5 (separate file)} - Structure 2 of the viscoelastic suspension flow: This video illustrates the flow structure 2 of a viscoelastic suspension decomposed by the optDMD method in real time.

\textbf{SI Video 6 (separate file)} - 30 vol\% particles assemble under shear: This video shows the assembly of 30 vol\% particles under shear in a viscoelastic suspension. Each pixel is 4.4 micrometers; the shear rate is 100/s; { the gap height is 0.2 mm}, and the video is 10x slower than the real-time. { Upon initiating the test, the motor of the rheometer induces a noticeable vibration in the image.}

\textbf{SI Video 7 (separate file).} - 5 vol percent suspension under LAOS cycle: This video presents the behavior of a 5 vol\% suspension under large amplitude oscillatory shear (LAOS) cycles. Each pixel is 4.4 micrometers, the strain amplitude is 3000\% and the oscillation frequency is 1 Hz. 

\textbf{SI Video 8 (separate file)} - 20 vol percent suspension under LAOS cycle: This video presents the behavior of a 20 vol\% suspension under large amplitude oscillatory shear (LAOS) cycles. Each pixel is 4.4 micrometers, the strain amplitude is 3000\% and the oscillation frequency is 1 Hz.

\textbf{SI Video 9 (separate file)} - 30 vol percent suspension under LAOS cycle: This video presents the behavior of a 30 vol\% suspension under large amplitude oscillatory shear (LAOS) cycles. Each pixel is 4.4 micrometers, the strain amplitude is 3000\% and the oscillation frequency is 1 Hz.

\textbf{SI Video 10 (separate file)} - polydisperse suspension under LAOS cycle: This video presents the behavior of a polydisperse suspension subjected to large amplitude oscillatory shear (LAOS) cycles. Each pixel is 4.4 micrometers, the strain amplitude is 3000\% and the oscillation frequency is 1 Hz.

\begin{figure*}
\centering
\includegraphics[width=3 in]{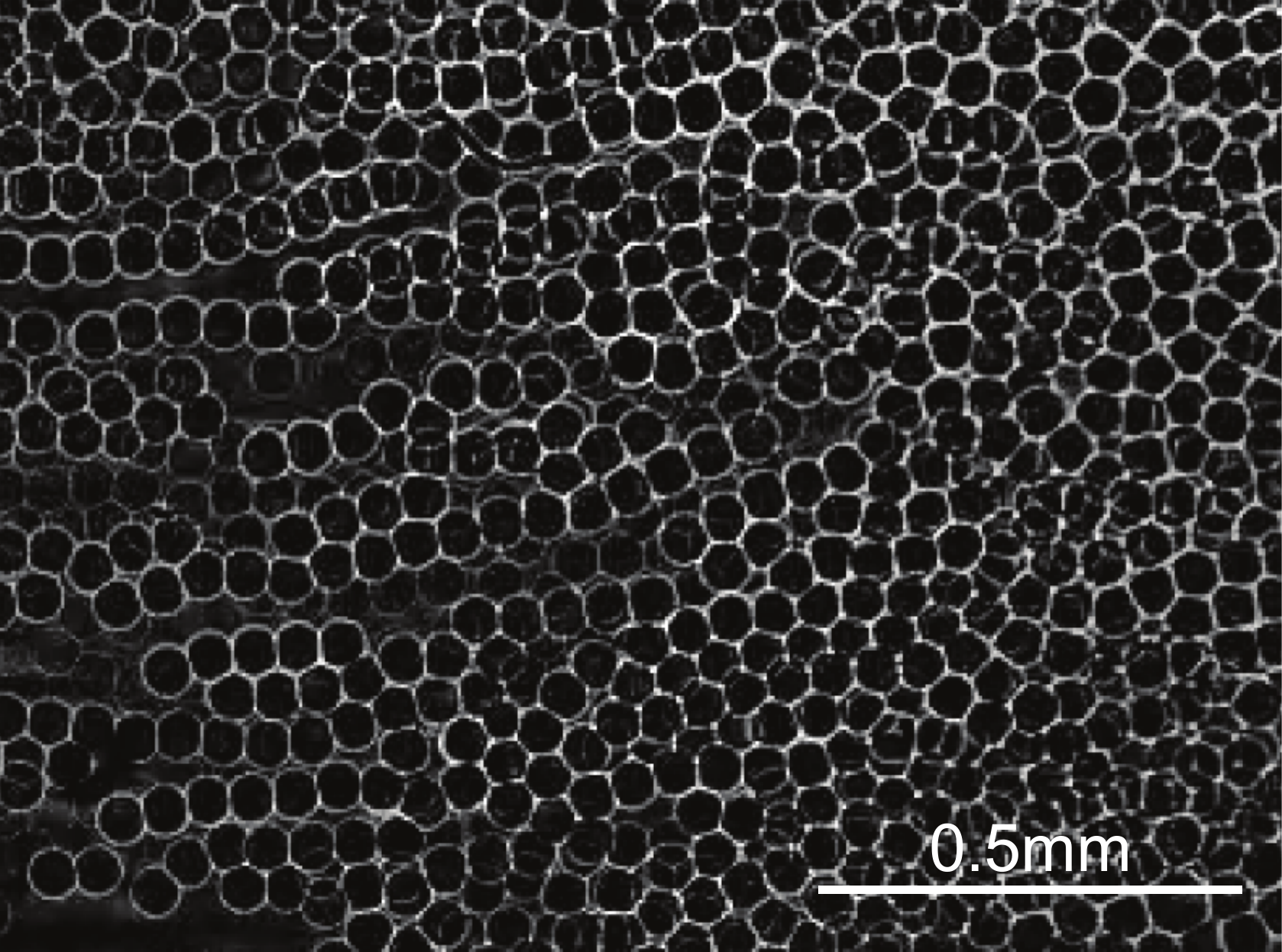}
\caption{The formation of suspended particles after around 2 seconds shearing at a shear rate $= 100\si{s^{-1}}$. The particles begin to {assemble into crystals. And the flow is unstable under this condition} .}
\label{fig:A3}
\end{figure*}





\begin{figure*}
\centering
\includegraphics[width=7 in]{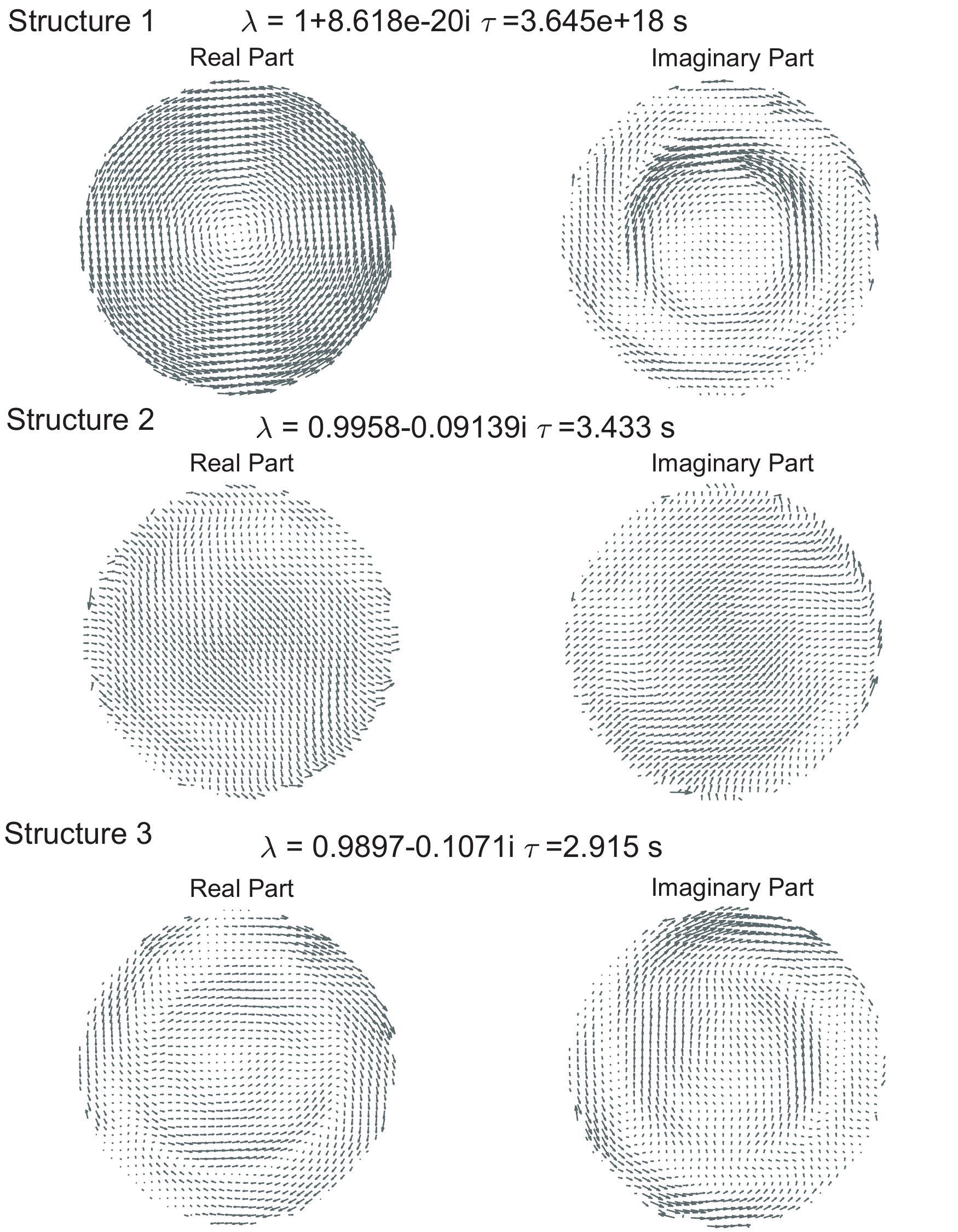}
\caption[The first three coherent structures of viscoelastic flow with the polydispersed suspension]{\label{fig:a7} The first three coherent structures of viscoelastic flow with the polydispersed suspension. $\lambda$ here is the associated eigenvalue and ~$\tau$  ~is the oscillatory period. The Left is the real part of the structure, and the right side is the imaginary part of the structure.} 

\end{figure*}